\documentclass[a4paper]{article}

\usepackage{INTERSPEECH2018}

\title{I-vector Transformation Using Conditional Generative Adversarial Networks for Short Utterance Speaker Verification}
\name{Jiacen Zhang, Nakamasa Inoue, Koichi Shinoda}
\address{
  Tokyo Institute of Technology}
\email{\{jiacen, inoue\}@ks.c.titech.ac.jp, shinoda@c.titech.ac.jp}

\begin{document}

\maketitle

\begin{abstract}
  I-vector based text-independent speaker verification (SV) systems often have poor performance with short utterances, as the biased phonetic distribution in a short utterance makes the extracted i-vector unreliable. 
  This paper proposes an i-vector compensation method using a generative adversarial network (GAN), where its generator network is trained to generate a compensated i-vector from a short-utterance i-vector and its discriminator network is trained to determine whether an i-vector is generated by the generator or the one extracted from a long utterance. Additionally, we assign two other learning tasks to the GAN to stabilize its training and to make the generated i-vector more speaker-specific. Speaker verification experiments on the NIST SRE 2008 ``10sec-10sec'' condition show that our method reduced the equal error rate by 11.3\% from the conventional i-vector and PLDA system.
    
  
\end{abstract}

\section{Introduction}

Recent years have seen a great improvement in text-independent speaker verification. The speaker verification system extracts speaker characteristic information from a given utterance and then verify the speaker ID. In the state-of-the-art methods of speaker verification, i-vector \cite{i-vector} is used to represent speaker characteristics, and probabilistic linear discriminant analysis (PLDA) \cite{plda0,plda1,plda2} is used as a verifier. While this system performs well on long utterances, the performance degrades drastically when only short utterances are available \cite{kanagasundaram2011vector}. The main cause of this problem is the biased phonetic distribution of short utterances, which makes the estimated speaker features become statistically unreliable. However, in many real world scenarios, users may be reluctant to provide several-minute-long utterances. 

Significant efforts have been made to remedy the performance degradation in short utterance speaker verification. In \cite{kanagasundaram2014improving}\cite{7953206}\cite{duration}, the variance of i-vectors for short utterances are modeled and used for i-vector normalization. \cite{kenny2013plda} and \cite{lin2017fast} proposed to utilize duration information in PLDA model. \cite{yamamoto2015denoising} uses phonetic information to reconstruct reliable i-vectors.

In the past years, deep learning has become very popular in the speaker verification field. Many approaches use deep neural networks to process i-vectors. For example,  \cite{villalba2017tied} proposed a variational autoencoder as a back-end for i-vector based speaker recognition, \cite{mahto2017vector} used denoising autoencoders to compensate for noisy speech. However, a large amount of data is required for training deep neural networks \cite{e2e}, while the amount of data available for speaker verification are usually very small. This has been one of the biggest obstacles for building an end-to-end speaker verification system using deep learning. Hence, it may be better to improve the i-vector and PLDA framework by using deep learning.
Recently, a novel structure called generative adversarial network (GAN) \cite{goodfellow2014generative} has become extremely popular. GAN can learn a mapping from random noise to target domain, by playing a zero-sum game with two networks, a generator $G$ and a discriminator $D$: $G$ tries to generate ``real'' samples which can fool $D$, while $D$ tries to determine whether a given sample is from real data distribution or from $G$.

This paper describes an i-vector transformation method using conditional GAN for improving i-vector based short utterance speaker verification. The method uses GAN to estimate a generative model which can generate a reliable i-vector from an unreliable i-vector, in which we assume an i-vector from a long utterance is reliable, and an i-vector from a short utterance is unreliable. Specifically, we used the conditional version of GAN, where both the generator and the discriminator have an i-vector from a short utterance as the conditional input. The generator $G$ tries to generate a reliable i-vector from an unreliable one, and the discriminator $D$ tries to decide whether a given reliable i-vector is a real one extracted from a long utterance or a fake one generated by $G$. In order to stabilize GAN training, numerical difference (cosine distance) between generated i-vectors and target reliable i-vectors are used in the training stage. Moreover, inspired by \cite{mahto2017vector}, we tried to improve the speaker discriminative ability of generated i-vectors by adding an extra speaker label predicting task to $G$. This multi-task learning framework can better guide the training of GAN. In the testing stage, $G$ is used to generate reliable i-vectors from those extracted from short utterances, and then the generated i-vectors would be used in PLDA scoring.


This paper is organized as follows: Section 2 briefly introduces related works of our methods. Section 3 presents the proposed GAN-based structure for i-vector restoration. Section 4 describes experimental evaluations for speaker verification in two NIST SRE tasks. Section 5 summarizes this paper.

\section{Related Works}

\subsection{I-vector and PLDA}
I-vector and PLDA have been widely used in the state-of-the-art systems for text-independent speaker verification. 
The i-vector approach aims to extract a fixed and low dimension representation from a given utterance based on a factor analysis model. As described in \cite{i-vector}, an utterance is projected onto a low-dimensional total variability space which contains both channel- and speaker-dependent information, as an i-vector. Given an utterance, the channel- and speaker-dependent GMM supervector $M$ can be written as:
\begin{equation}
M = m + Tw,
\label{eq-2}
\end{equation}
where $m$ is the speaker- and channel-independent supervector taken from the universal background model (UBM), $T$ is the total variability matrix (TVM) and $w$ is the i-vector.

Probability linear discriminant analysis (PLDA) \cite{plda2} is applied as a generative model for i-vectors, which can be written as follows,
\begin{equation}
w = \bar{w} + Ux + Vy + \epsilon
\label{eq-1}
\end{equation}
where $\bar{w}$ is the global mean of i-vectors, $U$ and $V$ are an eigenvoice and an eigenchannel matrix respectively, $x$ and $y$ are speaker- and channel-factors, and $\epsilon$ is residual noise.

Given two i-vectors, the log-likelihood ratio of the same-speaker and different-speaker hypotheses is computed by the PLDA model as the measure of their similarity. 

\subsection{Generative Adversarial Networks Family}

Generative adversarial networks (GANs) were introduced in \cite{goodfellow2014generative} to estimate a generative model by an adversarial process, in which a generator G tries to generate a sample using a random noise vector $z$ and a discriminator D tries to compute the probability that a given sample is from real data $y$ rather than generated by G. Training of GAN is equivalent to optimizing the following min-max function,
\begin{equation}
\begin{aligned}
\min\limits_G \max\limits_D V_{\textnormal {GAN}}(D, G) =&\ E_{y}[\log D(y)] \\ 
                            &+ E_{z}[\log (1-D(G(z)))].
\label{eq1}
\end{aligned}
\end{equation}
\begin{figure}[t]
  \centering
  \includegraphics[scale=0.5]{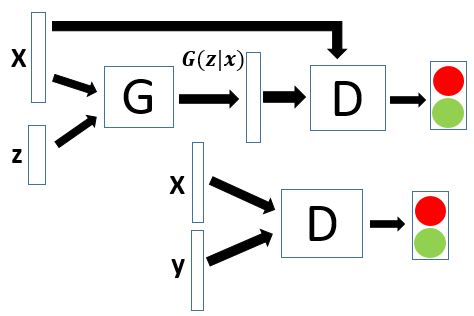}
  \caption{Conditional Generative Adversarial Networks.}
  \label{fig:conditional GAN}
\end{figure}
As our target is about transformation, we used GAN's conditional version (CGAN) \cite{mirza2014conditional} in our approach. The adversarial training procedure is almost the same as the original GAN, and the only difference is both the generator and the discriminator have a conditional input $x$, as in Figure 1. The min-max function is:
\begin{equation}
\begin{aligned}
\min\limits_G \max\limits_D V_{\textnormal {CGAN}}(D, G)=\  & E_{x,y}[\log D(y|x)]\\
           &+ E_{x,z}[\log (1-D(G(z|x)))].
\label{eq2}
\end{aligned}
\end{equation}
There have already been several successful applications of CGAN in similar tasks. \cite{isola2017image} uses it to convert image styles. \cite{pascual2017segan} applies it to enhance speech. Inspired by their success, we apply CGAN in i-vector space to improve the performance of i-vector based short utterance speaker verification.  

\section{Proposed Method}
\begin{figure}[t]
  \centering
  \includegraphics[scale=0.4]{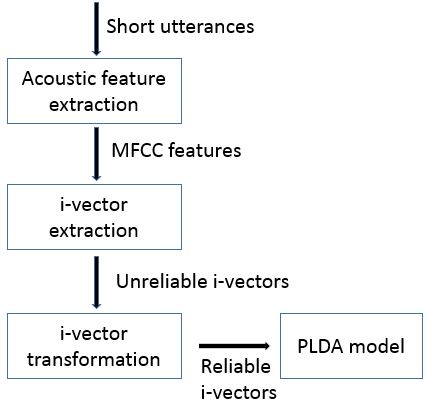}
  \caption{Framework of our method.}
  \label{fig:general framework}
\end{figure}
Figure 2 shows the framework of our proposed method. At first, acoustic features (MFCC) are extracted from a short utterance, then an unreliable i-vector is extracted from them. Next, an i-vector transformation function is applied to the unreliable i-vector, and finally the transformed i-vector is fed into the PLDA model.
 
\subsection{GAN for i-vector Transformation}
Our target is to estimate a transformation function which can restore a reliable i-vector (extracted from a long utterance), from a short-utterance i-vector. We use a CGAN-based structure to estimate this function. Overall architecture of the proposed GAN is the one shown in Figure 1, where the conditional input $x$ is an i-vector extracted from a short utterance, the real sample $y$ is an i-vector from a long utterance. In the training stage, $G$ is optimized to generate a reliable i-vector using the one extracted from a short utterance, and $D$ is optimized to determine whether the given reliable i-vector is fake (generated by $G$) or real (extracted from a long utterance). In testing, $G$ is used as the transformation function for an i-vector extracted from a short utterance in the testing set.

In order to prevent several problems such as unstable gradient and model collapse in GAN training, we use a special GAN structure Wasserstein GAN (WGAN) \cite{arjovsky2017wasserstein}. Denoting $x$ as an unreliable i-vector, $y$ as a reliable i-vector and $z$ as random noise, the min-max function is represented as:
\begin{equation}
\begin{aligned}
\min\limits_G \max\limits_D V_{\textnormal {WCGAN}}(D, G) =\ &E_{x,y}D(y|x) \\
& - E_{x,z}D(G(z|x)),
\label{eq3}
\end{aligned}
\end{equation}
Then the objective function related to GAN for $G$ is 
\begin{equation}
\min \textnormal {G}= - E_{x,z}D(G(z|x)),
\label{eq4}
\end{equation}
and for $D$, objective function is
\begin{equation}
\max \textnormal{D} = E_{x,y}D(y|x) - E_{x,z}D\left(G\left(z|x\right)\right).
\label{eq5}
\end{equation}
Regarding the training data for GAN, i-vectors extracted from short and long utterances are required. While only long utterances are present in the training dataset, we obtained short utterances by segmenting a long utterance into short utterances. I-vectors are extracted from both long and short utterances using the same extractor. Through this process we can obtain an i-vector pair consisting two i-vectors from the same speaker and session, but one is from a short utterance and the other is from a long utterance. The i-vector pairs are utilized in the next section.
   
\subsection{Speaker Verification-oriented Objective Functions}
To better guide the training of GAN for our task and make the best use of the training data, two additional learning tasks are added to the GAN framework. 

\subsubsection{Numerical difference}
The most straight-forward approach to measure the performance of transformation is computing the numerical difference between the generated i-vector and the target. We compute this objective function using i-vector pairs mentioned above. In many other similar tasks, mean squared error (MSE) is used to measure such a numerical difference. However, for i-vectors, we believe cosine distance is more suitable. The objective function related to this task can be written as:
\begin{equation}
\min \textnormal {COS} = \frac{1}{m}  \sum_{i=1}^{m} \left[\frac{1}{n_i} \sum_{j=1}^{n_i}\left(1 - \frac{G(z|x_{ij}) \cdot y_i}{\|G(z|x_{ij})\|\ \|y_i\|} \right)\right],
\label{eq6}
\end{equation}
where $m$ is the number of long utterances in the training set, $y_i$ refers to the i-vector extracted from the $i$-th long utterance in the training set, $n_i$ is the number of short utterances extracted from the $i$-th long utterance, $x_{ij}$ means the i-vector extracted from the $j$-th segment of the $i$-th long utterance and $z$ is random noise.
\begin{figure}[t]
  \centering
  \includegraphics[scale=0.45]{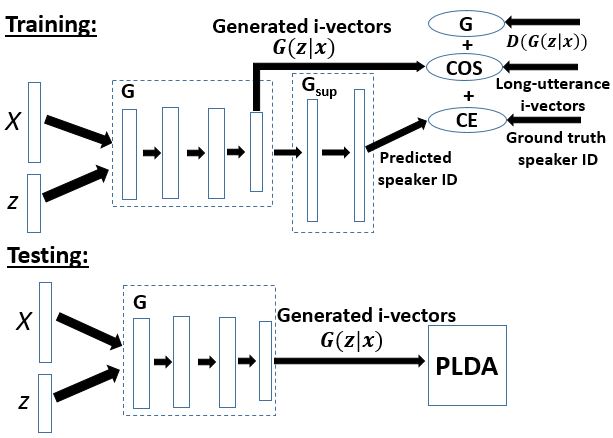}
  \caption{Training of the generator network $G$ and its application in the testing stage.}
  \label{fig:modified G}
\end{figure}
\subsubsection{Speaker discrimination}
The training objectives explained above only compensate the variance brought by the biased phonetic distribution of short utterances, and the speaker labels provided by the training set are not used yet. 
Clearly, improving the speaker discriminative ability of generated i-vectors can enlarge the inter-speaker differences among i-vectors, which would improve the verification performance in the PLDA scoring stage. As shown in Figure 3, in the training stage, a supplementary section, $G_{\textnormal{sup}}$, is concatenated after the generator $G$, which takes the generated i-vector as an input and predicts its speaker label. We minimize cross entropy between the prediction result and the ground truth:
\begin{equation}
\min \textnormal{CE} = \frac{1}{m}\sum_{i=1}^{m} \left[\frac{1}{n_i} \sum_{j=1}^{n_i} l_{ij}^k \left(\log o_{ij}^k\right)\right],
\label{eq7}
\end{equation}
where $l_{ij}^k$ is the empirical probability observed in the ground truth that the target i-vector belongs to the $k$-th class, and $o_{ij}^k$ is the predicted probability that the generated i-vector belongs to the $k$-th class. In summary, for training $G$, our goal is to minimize
\begin{equation}
a\ \textnormal{G} + b\ \textnormal{COS} + c\ \textnormal{CE},
\label{eq8}
\end{equation}
where $a$, $b$, $c$ are weight parameters for these three targets, respectively.

After training, as shown in Figure 3, only $G$ is used to generate a reliable i-vector, which is fed into a PLDA model for the next scoring step.

\section{Evaluation}

\subsection{Experimental setup}
We evaluated the performance of our method in the speaker verification tasks of the NIST SRE 2008 \cite{sre08}. We used the ``short2-10sec" and ``10sec-10sec" conditions as our trial sets, where each session is an excerpt of telephone speech, and ``short2" refers to five-minute-long speech while ``10sec" means that the active voice part in the sample is about 10 seconds. There are three sub-conditions in the trail sets: Condition 6 covers all the speech segments, Condition 7 involves only those spoken in English, and Condition 8 only has those spoken in English by native U.S. English speakers \cite{sre08}. 
 Performance measures for the evaluation were the equal error rate (EER) and the minimum detection cost function (minDCF) of NIST SRE 2008 \cite{sre08} on the trails calculated with DETware provided by NIST \cite{det}. 

We compared our method, which is named as ``D-WCGAN'' (Discriminative WCGAN) in the experiments with a baseline i-vector and PLDA system that does not apply any short-utterance compensation techniques. To demonstrate the contribution of GAN to the performance improvement, we made an extra system, which shared almost the same structure with the proposed GAN but did not contain a discriminator and did not use GAN-related objective function. This system is named as ``Single G'' in the following part. 

\subsubsection{Baseline system}
The baseline system is the i-vector and PLDA system shown in Section 2. In this system, the input speech segment was first converted to a time series of 60 dimensional feature vectors of Mel-frequency cepstral coefficients (20 dimensional features followed by their first and second derivatives) extracted from a frame of 20ms long and 10ms shift. An i-vector of 400 dimensions was then extracted from the acoustic features using a Gaussian mixture model with 2048 mixture components as a universal background model (UBM) and a total variability matrix (TVM). Length normalization was applied to i-vectors as a preprocessing step before being sent to the PLDA model. Kaldi speech recognition toolkit \cite{kaldi} was used to run these steps.

The UBM, the TVM, and PLDA models were all gender-dependent and trained with SRE08's development data, which contains the NIST SRE2004-2006 data, Switchboard, and Fisher corpus. This dataset as a whole consistes 34,925 utterances from 7,275 male speakers.
\subsubsection{Proposed GAN}
The training data of GAN is a subset of SRE08's development set mentioned above and SRE08's training set, which contains 1,986 male speakers in total. To make the short and long utterance pairs mentioned in Section 4, we used a sliding window of 20s long and 10s shift to cut one long utterance into short utterances. The UBM, TVM for extracting i-vectors are the same as the one used in the baseline system. Finally, we got 331,675 i-vector pairs for GAN training.
The activation function of hidden layers in the proposed GAN, if not specified, is a leaky ReLU \cite{lrelu} with an alpha value set to 0.3. As mentioned above, $G$ generates an i-vector and $G_{\textnormal{sup}}$ predicts its speaker label. The input layer of $G$ contains 450 nodes to accept the 400-dimension i-vectors and random noise vectors of 50 dimensions, followed by three hidden layers with 512 nodes. $G$'s output layer has 400 nodes, which holds the generated i-vector. The activation function for the output layer of $G$ is tanh. $G_{\textnormal{sup}}$ has one hidden layer, which contains 1,986 nodes. Output layer of $G_{\textnormal{sup}}$ also have 1,986 nodes and the activation function of each node is softmax.The random noise vectors were sampled from a Gaussian distribution with zero mean and standard deviation $0.5$. $D$ has four hidden layers and its input layer has 800 nodes, which accepts two concatenated i-vectors. Output layer of $D$ has only one node with a linear activation function. As we used the WGAN structure, weight clipping is done on $D$, where the clipping range is $-0.01$ to $0.01$. 

We used the Tensorflow library \cite{tensorflow} for our neural networks implementation. The networks were optimized using RMSProp \cite{rmsprop} with a mini-batch of 64 samples. The learning rate was set to $0.0001$. For G training, we set the value of $a$, $b$, $c$ as 4, 7, 1, respectively.

In the testing phase, for the ``short2-10sec" condition, an i-vector extracted from an utterance in the testing set are transformed by $G$, then PLDA scoring is done on the i-vector extracted from the enrollment set and the transformed i-vectors. At last, score-wise fusion is done between the baseline system and the proposed method. For the ``10sec-10sec" case, almost all the steps are the same as the former one, but the i-vectors from both the enrollment and the testing set are transformed by $G$. The i-vector extractor and PLDA model are the same as those used in the baseline system.

\begin{table}[t]
\caption{The speaker verification results in terms of EER (\%) on all the three conditions of the SRE08 ``short2-10sec" male trail list.}
  \label{tab:example}
\centering
\begin{tabular}{l@{} r r r r}
\toprule
\multicolumn{1}{c}{}&
\multicolumn{4}{c}{\textbf{EER (\%)}} \\
\cmidrule{2-5}
\multicolumn{1}{c}{\textbf{System}} & 
\multicolumn{1}{c}{\textbf{Cond. 6}} &
\multicolumn{1}{c}{\textbf{Cond. 7}}  &
\multicolumn{1}{c}{\textbf{Cond. 8}} & 
\multicolumn{1}{c}{\textbf{Average}} 
\\
\midrule
a) Baseline   & 7.28 & 6.15 & 6.06 & 6.50 \\
b) Single G   & 10.04 & 8.85 & 8.33 & 9.07 \\
c) a + b      & 7.28 & 5.77 & 6.06 & 6.37 \\
d) D-WCGAN    & 9.45 & 8.08 & 8.33 & 8.62 \\
e) a + d      & \textbf{6.89} & \textbf{5.77} & \textbf{5.30} & \textbf{5.99} \\
\bottomrule
\end{tabular}
\end{table}

\begin{table}[t]
\caption{The speaker verification results in terms of EER (\%) on all the three conditions of the SRE08 ``10sec-10sec" male trail list.}
  \label{tab:example}
\centering
\begin{tabular}{l@{} r r r r}
\toprule
\multicolumn{1}{c}{}&
\multicolumn{4}{c}{\textbf{EER (\%)}} \\
\cmidrule{2-5}
\multicolumn{1}{c}{\textbf{System}} & 
\multicolumn{1}{c}{\textbf{Cond. 6}} &
\multicolumn{1}{c}{\textbf{Cond. 7}}  &
\multicolumn{1}{c}{\textbf{Cond. 8}} & 
\multicolumn{1}{c}{\textbf{Average}} 
\\
\midrule
a) Baseline   & 11.97 & 10.32 & 9.60 & 10.63 \\
b) Single G   & 15.32 & 13.89 & 12.00 & 13.77 \\
c) a + b      & 11.16 & 10.71 & 9.60 & 10.49 \\
d) D-WCGAN     & 15.42 & 13.89 & 13.60 & 14.30 \\
e) a + d      & \textbf{10.75} & \textbf{8.73} & \textbf{8.80} & \textbf{9.43} \\
\bottomrule
\end{tabular}
\end{table}

\begin{table}[t]
\caption{The speaker verification results in terms of minDCF on Condition 6 of the SRE08 ``short2-10sec" and ``10sec-10sec" male trail lists.}
  \label{tab:example}
\centering
\begin{tabular}{l@{} r r}
\toprule
\multicolumn{1}{c}{}&
\multicolumn{2}{c}{\textbf{minDCF}} \\
\cmidrule{2-3}
\multicolumn{1}{c}{\textbf{System}} & 
\multicolumn{1}{c}{\textbf{short2-10sec}} &
\multicolumn{1}{c}{\textbf{10sec-10sec}} 
\\
\midrule
a) Baseline   & \textbf{0.370} & 0.553 \\
b) Single G   & 0.494 & 0.717 \\
c) a + b      & 0.391 & 0.540  \\
d) D-WCGAN    & 0.454 & 0.678 \\
e) a + d      & 0.375 & \textbf{0.522}  \\
\bottomrule
\end{tabular}
\end{table}

\subsection{Results}
 Table 1 shows the EERs of the ``short2-10sec" condition of NIST SRE 2008. The average EER of our proposed method was 5.99\%, and it outperformed that of the baseline i-vector PLDA system, 6.50\%. The reduction of average EER is 7.85\%. Table 2 shows the EERs of ``10sec-10sec" condition of NIST SRE 2008. The average EER of our proposed method was 9.43\%, and it outperformed the 10.63\% of baseline, and the reduction of average EER is 11.29\%. Although our method alone did not outperform the baseline system, it achieved better results when the score-wise fusion was done with the baseline method. We found that the best results was achieved when the score weight ratio of baseline system and our method is 7:3. Table 3 shows the minDCF of the Condition 6 of ``short2-10sec" and ``10sec-10sec" sets. The minDCF of our method is 1.33\% worse than the baseline's in ``short2-10sec", but 5.61\% better in ``10sec-10sec". These results showed that our proposed method can make i-vectors more reliable in most cases. However, in current stage, the amount of training data for the GAN is not enough, even smaller than the amount of PLDA's training data. If we have more training data for the GAN, the performance of the proposed methods may become much better.

Regarding the importance of GAN, our results (b, c in Table 1, 2 and 3) showed that performance became worse, but slightly better than the baseline system in EER, when $D$ was absent. This fact demonstrates the contribution of GAN.

\section{Conclusions}
This paper has proposed a GAN-based speaker feature restoration method for speaker verification using short utterances. The generator is trained to transform an unreliable i-vector extracted from a short utterance to a reliable i-vector which can be extracted from a long utterance. Speaker labels are also used in the training of GAN to improve the speaker discriminative ability of generated i-vectors. The evaluation results on NIST SRE 2008 task show that our proposed method improved the performance, especially when only short utterances are available for enrollment and testing. 

Our future work includes collecting more data for GAN training, as well as applying the GAN-based framework to other cases when i-vectors become unreliable, for example, noise exists in utterances. In addition, we plan to make the discriminator network able to determine whether two given i-vectors are from one speaker or not, so that we can use the GAN model as a back-end for the text-independent speaker verification system. 


\section{Acknowledgment}
This work was supported by JSPS KAKENHI 16H02845 and by JST CREST Grant Number JPMJCR1687, Japan.
\bibliographystyle{IEEEtran}

\bibliography{mybib}

\end{document}